\documentstyle[preprint,floats,aps,epsf,epsfig,prb]{revtex}

\newcommand \ltdash{\raise-1.8pt\hbox{$\scriptscriptstyle |$}}

\newcommand \bea {\begin{eqnarray} }
\newcommand \eea {\end{eqnarray}}

\begin{document}
\draft
\title{
Breakdown of the Chiral Luttinger Liquid in One-Dimension. 
}
\author{ A. F. Ho$^{1}$ and  P. Coleman$^{1}$ }
\address{
$^1$Serin Laboratory, Rutgers University, P.O. Box 849,
Piscataway, New Jersey 08855-0849}
\maketitle
\date{\today}
\maketitle
\begin{abstract}
We have developed a fermionic boot-strap method to solve a  class of
chiral one dimensional  fermion model  which  cannot be tackled by
bosonization. Using this scheme, we show that Luttinger liquid
behavior in a  gas of four interacting chiral Majorana fermions is
highly sensitive to the velocity degeneracy.  Upon  
 changing  the velocity of a one chiral fermion,
a 
 sharp bound (or anti-bound) state  splits off
from the original Luttinger liquid continuum,
cutting off the X-ray singularity to form a broad incoherent excitation
with a lifetime that grows linearly
with frequency. 

\end{abstract}

\vskip 0.2 truein
\pacs{}

\section{INTRODUCTION}

The anomalous normal state behavior discovered in the cuprate 
superconductors has stimulated enormous interest in the possibility
of new kinds of electronic fluid that might provide an alternative
to Fermi liquid behavior.
The classic model for non-Fermi liquid behavior is provided by
the one-dimensional electron gas, 
where  the generic fixed point behavior is a 
a Luttinger  Liquid.\cite{Haldane} 
Luttinger liquid behavior in one dimension
is associated with two special features:  a 
the Fermi surface which consists
of just two points $\pm k_f$ where the electrons 
interact very strongly  and second, the special kinematics
of one dimension, whereby
energy  and momentum conservation  impose a {\sl single}
constraint on scattering processes near the Fermi surface, 
which gives rise to a qualitative
enhancement in scattering phase
space. 
The combination of these two factors cause the electron to lose its  
eigen-state status to the  collective spin and charge density bosonic modes.

In this paper, we introduce a generalization of the Tomonaga
Luttinger model, written
\bea
H = \int dx \left\{ -i \sum_{a=0}^3 v_a \Psi^{(a)}(x) \partial_x \Psi^{(a)}(x) 
+ g \Psi^{(0)}(x) \Psi^{(1)}(x) \Psi^{(2)}(x) \Psi^{(3)}(x) \right\}   ,
\eea
where the $\Psi^{(a)}$ ($a=(0,1,2,3)$ ) represent four real (Majorana)  
fermions such that 
$\Psi^{(a)}(x)=\Psi^{(a) \dagger}(x)$. The fermions are chiral 
(right-movers,say): this  is a 
crucial property that ensures the system stays gapless, and allows for exact solutions in
a number of cases. In the special case where
the velocities of each of these modes are degenerate, this model has an
O(4) symmetry, and 
corresponds to the chiral Luttinger liquid. In this case, the four Majorana
modes can loosely be associated with the spin up and spin down electron
and hole excitations of the Fermi surface.  

We shall show that
once we break the degeneracy
of the velocities by changing the velocity of a single
Majorana model,  a qualitatively new type of behavior develops.
The model  where only 
three of the velocities are equal, so that $v_{1,2,3}= v\ne v_0$, has  an
$O(3)$ symmetry.
This model  is actually  motivated from two rather disparate sources:
\begin{itemize}
\item It is inspired
in part by the transport phenomenology of the 
cuprates\cite{2tau}, which suggests that electrons near the
Fermi surface might divide up into two Majorana modes with 
different scattering rates and dispersion. To date,
this kind of behavior has only been realized in 
impurity models, and their infinite dimensional generalization.\cite{2CCK,HoColeman}
We shall show that by breaking the
velocity degeneracy of the original chiral Luttinger model, we obtain
a one-dimensional realization of this behavior, 
where a sharp Majorana mode
intimately co-exists with an incoherent continuum of excitations, reminiscent
of the higher dimensional phenomenology. 

\item Frahm et. al.\cite{Frahm} have recently proposed that the low energy
Hamiltonian of an integrable spin-1 Heisenberg chain 
doped with mobile spin-$1/2$ holes is 
given by Eqn.1, with one Majorana fermion $\Psi^{(0)}$ describing a slow moving
excitation coming from the dopant, interacting with three
rapidly-moving Majorana fermions that describe the spin-1 excitations
of the spin-chain. Such doped
spin-chain models may be relevant to certain experimental systems such as
$Y_{2-x} Ca_x Ba Ni O_5$.\cite{spinchain}
\end{itemize} 
\begin{figure}[tb]
\epsfxsize=5truein
\centerline{\epsfbox{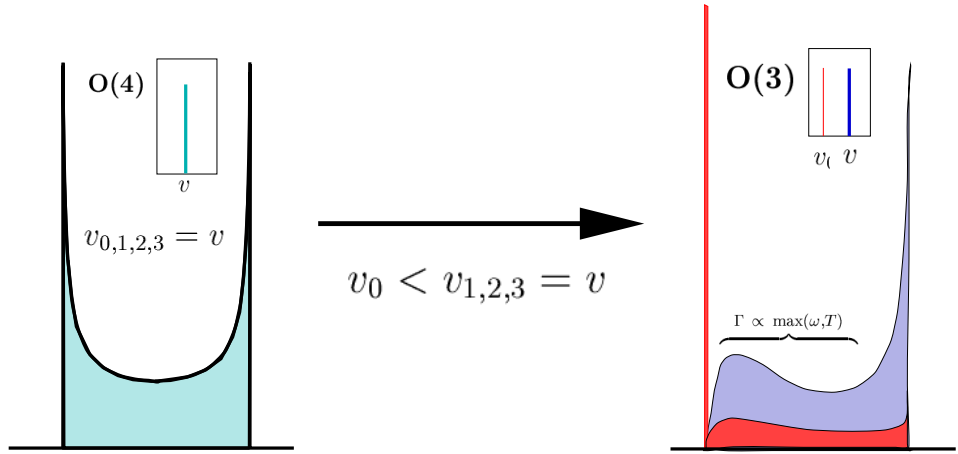}}
\vskip 0.3 truein
\caption{Schematic diagram showing the evolution of the spectral
weight as we introduce velocity difference to the fermions.
Inset indicates the bare spectral function, without interactions.}
\end{figure}
 
Whereas the  O(4) model can be treated
by bosonization,\cite{Haldane,Voit} 
changing the velocity of a single Majorana fermion introduces
non-linear terms into the  bosonized Hamiltonian that preclude
a separation in terms of Gaussian  spin and charge degrees of freedom. 
We shall show, using a new fermionic approach to the problem,  that
when  we break the degeneracy
of the velocities , the X-ray catastrophe
associated with the Luttinger liquid behavior is cut-off by the
velocity shift. The ``horn-like'' feature in the spectral weight of
the Luttinger liquid is then split into a sharp bound (or anti-bound) state
that co-exists with an incoherent spin-charge decoupled continuum, with a
scattering rate linear in frequency.
We summarize these results in the schematic diagram Fig.1.


\section{METHOD:PHILOSOPHY}

We now develop a fermionic  scheme for solving both the
 $O(4)$ and the $O(3)$ versions of the chiral Majorana
model. 
Our method is based on the observation that for these models, the  skeleton self-energy, without vertex corrections,
is {\it exact}, so that 
\bea
\Sigma_{a}(x,\tau) = g^2 G_{b}(x,\tau) G_{c}(x,\tau) G_{d}(x,\tau) ,
\eea
where the $G_{a}$ are the Greens functions of the interacting
system and $\{a,b,c,d\}$ is
a cyclic permutation of $\{0,1,2,3\}$.
We  can represent this  result diagrammatically  in Fig.2.
These equations close with the usual relations:
\bea
\Sigma_{a}(k,\omega) = (i\omega - v_a k) - 
G_{a} (k,\omega)^{-1}  .  \qquad (a=0,1,2,3)
\eea
These
coupled equations (3,4) together define a boot-strap method to solve the problem. 
Our method bears marked similarity to the   Non-Crossing 
Approximation used in solving various magnetic impurity models\cite{Bickers}, 
but in this case
case, no approximation is involved. 

\begin{figure}[tb]
\unitlength1.0cm
\begin{center}
\begin{picture}(6,1.8)
\epsfxsize=6.0cm
\epsfbox{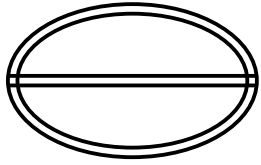}
\end{picture}
\vskip 0.3 truein
\caption{NCA self-energy. Double lines represent the exact propagator. }
\end{center}
\end{figure}

Provided that we have a minimal $O(3)$ symmetry, then the
three current densities    $j^a(x) = -i\epsilon_{abc} \Psi^{(b)}(x) \Psi^{(c)}(x)$ ($a,b \in (1,2,3)$) are conserved. Following 
Dzyaloshinskii and Larkin (1974)\cite{D&L,Metzner,Kopietz}, since
charge and current are the same in a chiral model, the continuity
equation guarantees that the 
N-point connected current-current correlation functions 
vanish  for $N>2$:
\bea
\langle j^a(1)j^a(2) \dots j^a(N) \rangle_C=0, \qquad \qquad (N>2)
\eea
For the non-interacting system,
this result leads to the ``loop cancellation theorem'', which means
that if we take the amplitude associated with 
a closed fermion loop with $N>2$ conserved current insertions, then 
the sum over all possible permutations $P$
of the space-time indices of the current operators must give  
zero\cite{D&L,Metzner,Kopietz} 
Dzyaloshinskii and Larkin used this fact to eliminate all diagrams that
contain such closed loops, considerably simplifying the calculation of the
vertex function and polarization bubbles.  

We use the loop cancelation theorem in a new way, to show that
the vertex  corrections to the skeleton self-energy identically vanish.
To illustrate the idea, consider the self-energy of the singlet 
Majorana mode in the O(3) model.  The NC contributions to its self-energy
are entirely constructed by combining loops with two current insertions,
as shown in Fig 3(iv). 
\begin{figure}[tb]
\unitlength1.0cm
\begin{center}
\begin{picture}(10,6)
\epsfxsize=10.0cm
\epsfbox{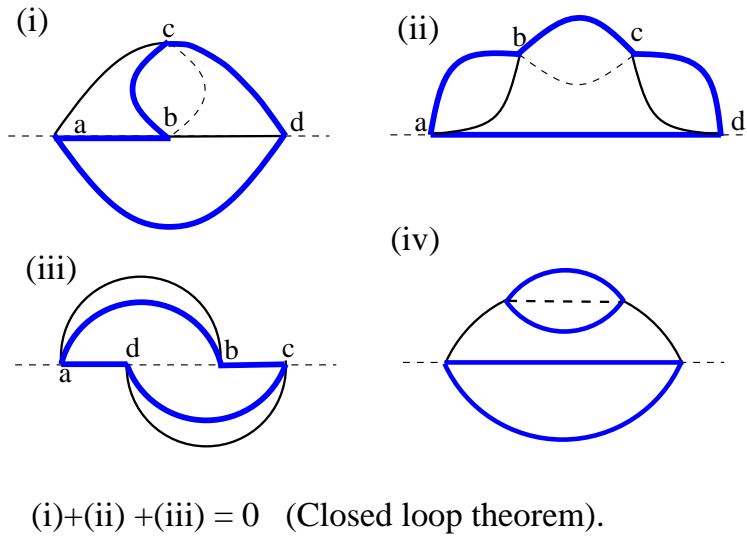}
\end{picture}
\vskip 0.3 truein
\caption{Order $g^4$ diagrams for the $O(3)$ model, demonstrating how the closed-loop
theorem works.  Dotted lines indicate bare propagator for  the singlet
Majorana fermion $\Psi^{(0)}$. Full lines indicate bare propagator
for the triplet Majorana fermions. Bold line highlights the loop 
with 2 or 4 current insertions.
}
\end{center}
\end{figure}
These contributions are finite. 
Non-skeleton diagrams involve
a sum of loops with more than two current insertions. For example, the
fourth order diagram, as shown in Fig. 3, involves a loop with
four current insertions, and these sum to give zero, 
leaving behind the single non-crossing contribution. This
result can be generalized to all higher order graphs, showing  that the
self-energy $\Sigma_0$ of the 
singlet Majorana fermion is given by the skeleton self-energy diagram without vertex
corrections.   
From this result,
we can show that the {\it full} Kadanoff-Baym Free energy  $F[G]$. The condition
that $\delta F[G]/\delta G=0$ generates the equations of motion for the self-energies.
But in order that this generate the skeleton self-energy diagram for $\Sigma_0$,
the Kadanoff-Baym Free energy Functional must 
{\sl truncate} at the leading skeleton diagrams.
\bea
F = - k_B T \{{\rm Tr}{\rm ln} [G^{-1}] - {\rm Tr}[ \Sigma G ] \}+
\parbox{3cm}{\epsfig{file=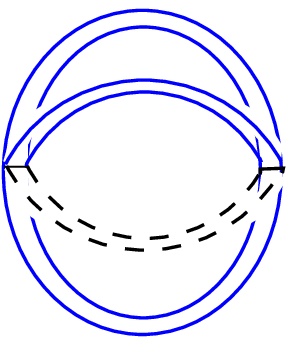,width=1cm,height=1.5cm}}
\eea
By differentiating the Free-energy functional with respect to the
Greens functions $G_{1,2,3}$  of the triplet Majorana fermions, 
we can then show that each
triplet self-energies are also given by the non-crossing skeleton expression. 

\section{METHOD-DETAILS}

We now apply the Non-Crossing approach (NCA) to the $O(3)$, using the limiting
case of the $O(4)$ model to check the validity of our results. 
Our equations are dramatically  simplified by seeking 
solutions to the NCA equations which satisfy a  scaling form
\bea
G_{a}(x,\tau) = \frac{1}{2 \pi i x}{\cal G}_{a}(\tau/ ix) ,
\eea
This  form is motivated by the observation that 
chirality prevents space from acquiring an anomalous dimension. 
Under a  Fourier Transform, this scaling function is self dual,
\bea
\frac{1}{2 \pi i x} {\cal G}_{a}(\tau /i x) 
\stackrel{F.T.}{\longleftrightarrow} 
\frac{1}{i \omega}{\cal G}_{a}(k/i\omega),
\eea
where the {\it same} function $ {\cal G}_{a}$ appears on both sides of the 
Fourier transform pair.
Inserting the scaling form into Eqn. (4)  and Fourier transforming the
resulting expression, we obtain
\bea
\Sigma_{a}(x,\tau) = - \frac{1}{2\pi i x} \frac{d^2}{du^2} 
\left[ 1-v_a u - 1/{\cal G}_{a} (u) \right]_{u=\tau/ix}  ,
\eea
Since the scaling form for the bare Green function is 
$1/{\cal G}_{a}^0 (u) = 1- v_a u $, it does
not contribute to the self-energy. Taking equation (3) together with equation (8),
we obtain the following differential equations:
\bea
\frac{d^2}{du^2} [{\cal G}_{a} (u)]^{-1} = 
- (g/2 \pi)^2 {\cal G}_{b} (u) {\cal G}_{c} (u) {\cal G}_{d} (u)  
\eea
where $\{a,b,c,d\}$ are cyclic permutations of $\{0,1,2,3\}$.  The boundary 
conditions are:
\bea
{\cal G}_{a} (0) = 1, \\
{\cal G}'_{a} (0) = v_a .
\eea
derived from the physical requirement that at high frequencies, the fermions
are free particles, moving with the {\it bare} velocity $v_a$. For the $O(4)$ model,
where ${\cal G}_a(u) \equiv {\cal G}(u)$ $(a=0,3)$, 
eqs. (9-11) reduce to a single differential equation, 
\bea
\frac{d^2}{du^2} [{\cal G} (u)]^{-1} = 
- (g/2 \pi)^2 [{\cal G} (u)  ]^3  
\eea
for which the solution is
\bea
G(x,\tau)= \frac{1}{2\pi i x} \left[1- v_+ \tau/ix \right]^{-1/2} 
\left[1-v_- \tau/ix \right]^{-1/2} 
\eea where $v_{\pm}= v \pm (g/2\pi)$ and $v$ is the bare velocity.
Identical results can be obtained by bosonization. 
This provides an important consistency check on our results,  confirming
that the non-crossing self-energy is exact for the Tomonaga Luttinger model.

\section{Results for $O(3)$ model.}

As is well known, in the $O(4)$ model, the spectral weight has two 
square-root singularity
at the renormalized velocities $v_+,v_-$ (x-ray edge catastrophe), 
with a continuum in between (Fig.4).\cite{Voit}
We now show that if $\Delta v = v-v_0$ is finite, the X-ray edge catastrophe
is partially eliminated. 
If  $v_0 < v$, we find that low velocity  ``Horn'', originally with velocity $v_-$,
develops a 
sharp bound-state pole in the singlet channel, and a broad incoherent excitation
in the triplet
channel with a lifetime growing linearly in energy.
If $v_0> v$, the high   velocity ``horn'' splits off a singlet  anti-bound-state
and the triplet channel develops a high-velocity incoherent excitation. 
(Fig.5).  The sharp bound-state in the singlet channel develops once a 
velocity difference is introduced, because energy  and momentum
conservation now provide distinct constraints to scattering (unlike in the $O(4)$ model),
leading to much less phase space for  $\Psi^{(0)}$ to decay into.

To see this, we must analyze the 
the more general coupled differential equations for the $O(3)$ case, which take the form
\bea
\frac{d^2}{du^2} {\cal G} _3^{-1} &=& - (g/2 \pi)^2 ({\cal G}_3)^2 {\cal G}_0,\cr
\frac{d^2}{du^2} {\cal G} _0^{-1} &=& - (g/2 \pi)^2 ({\cal G}_3)^3. \label{ncadiff}
\eea
A very convenient way to discuss these equations is to map them onto a
central force problem of a fictitious particle. 
If we write $\vec r = ( {\cal G} _3^{-1}, {\cal G} _0^{-1})$, 
$\vec F = - (g {\cal G}_3/{2 \pi})^2(
 {\cal G}_0,  {\cal G}_3)$,
we notice that these equations can be cast in the form 
$\ddot{{\bf r}} = {\bf F}$. By inspection, ${\bf r} \times{\bf F}=0$, 
so the force is radial and we immediately
deduce that the ``angular momentum'', 
${\bf r} \times {\dot{{\bf r}}} = -\Delta v$ is a constant. 
If we use polar 
co-ordinates, $({\cal G} _3^{-1},{\cal G} _0^{-1}) = r (\cos \theta,
\sin \theta )$ the equations for the Green-function resemble 
the motion of a Fictitious particle 
under the influence of an anisotropic central force: 
\bea
\ddot r - \frac{\Delta v^2 }{r^3}&= &- (g/2 \pi)^2 \frac{1}{r^3 \cos^3  \theta 
\sin \theta}\cr
r^2 \dot \theta &=& -\Delta v,
\eea
with boundary condition $r(0)= \sqrt{2}, \theta(0)= \pi/4$.
In these equations 
the  velocity difference $\Delta v= v- v_0 $ provides a the repulsive
centrifugal force. 
The ``particle'' starts out at ${\bf r}_0 = (1,1)$. 
So long as the ``particle''  falls directly into the origin,
both ${\cal G}_3$ and ${\cal G}_0$ diverge
with X-ray singularities.  However, once $\Delta v$ is finite,
the orbit no longer passes through the origin, thereby  eliminating
the associated X-ray singularity in the spectral function.
The quantity $\zeta= (v-v_0)/g$ plays the role of a coupling
constant, and approximate  analytic solutions are possible in the limiting cases
of small and large $\zeta$. 

Suppose first, that $\Delta v >0$. In this case, $\theta\rightarrow 0$ at some finite
``time'' $u= \tau_o$, at which $r=C$ and $\dot \theta = - \Delta v /C^2$.  For $u\sim u_0$,  
it follows that $(r,\theta) = (C, C \dot \theta (u-u_0))$, from which we can
read off the following asymptotic behavior 
\bea
{\cal G}_3(u)^{-1} & \sim & C   ,\\
{\cal G}_0(u)^{-1}& \sim & \frac{\Delta v \tau_0}{C}(1- u\tau_0^{-1}),
\eea 
so that \bea
{\cal G}_0 \sim Z/(1 - v_0^* u)\eea
where 
\bea
Z ={C }/{\Delta v \tau_o} \qquad ,  v_0^* = \tau_0^{-1}.
\eea
Thus when one Majorana fermion move faster than
the others, an anti-bound state with spectral weight $Z$, moving with velocity
$v_0^*$, splits off  above the continuum. 
For $\Delta v >> \frac{g}{2 \pi}$
interactions can be ignored, so
$v_0^*\rightarrow  v_0$, and $Z\rightarrow 1^-$.  For $\Delta v << \frac{g}{2 \pi}$,
the ``motion'' of the fictitious particle  emulates that of the $O(4)$ model until the
angle $\theta$ approaches zero. We may estimate $\tau_0$ and $C$ by setting
\bea
\pi/4 = \int_0^{\tau_0}\frac{ \Delta v}{\tilde r^2(u)}du,\ \  C \approx \tilde r(\tau_0)
\eea
where $\tilde r = [2 ( 1- v_+ u )(1-v_-u)]^{1/2}$ is the solution 
to the $O(4)$ case.
This estimate gives
\bea
v_0^*  &=&v_+ + \frac{g  }{\pi}e^{-\frac{g}{2  \Delta  v }},\cr
Z &=&\left( \frac{\sqrt{2} g}{\pi \Delta v }\right)
e^{-\frac{g}{4  \Delta  v }}
\eea
indicating that the formation of the sharp anti-bound-state is non-perturbative
in the velocity difference. These results can be generalized to the case
where $\Delta v <0$ by replacing $v_+\rightarrow v_-$, $g\rightarrow -g$.

We have carried out numerical solutions of the differential equations (\ref{ncadiff})
for intermediate values of the coupling constant $\zeta$.  
The equations were integrated from the initial boundary conditions
using a standard adaptive integration routine. 
These results are
summarized  in Fig. 5. and Fig.6.


\begin{figure}
\unitlength1.0cm
\begin{center}
\begin{picture}(7,6)
\epsfxsize=7.0cm
\epsfbox{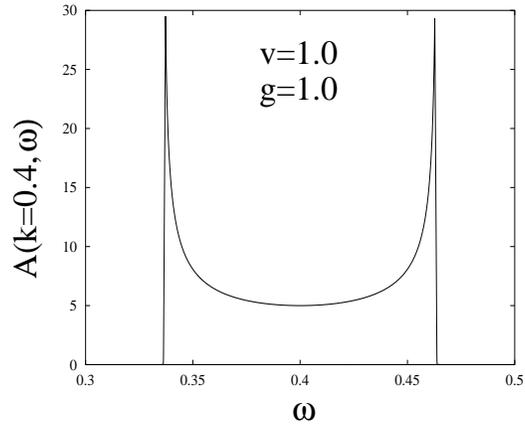}
\end{picture}
\vskip 0.4 truein
\caption{Spectral weight of O(4) model}
\end{center}
\end{figure}

\begin{figure}
\unitlength1.0cm
\begin{center}
\begin{picture}(12,8)
\epsfxsize=12.0cm
\epsfbox{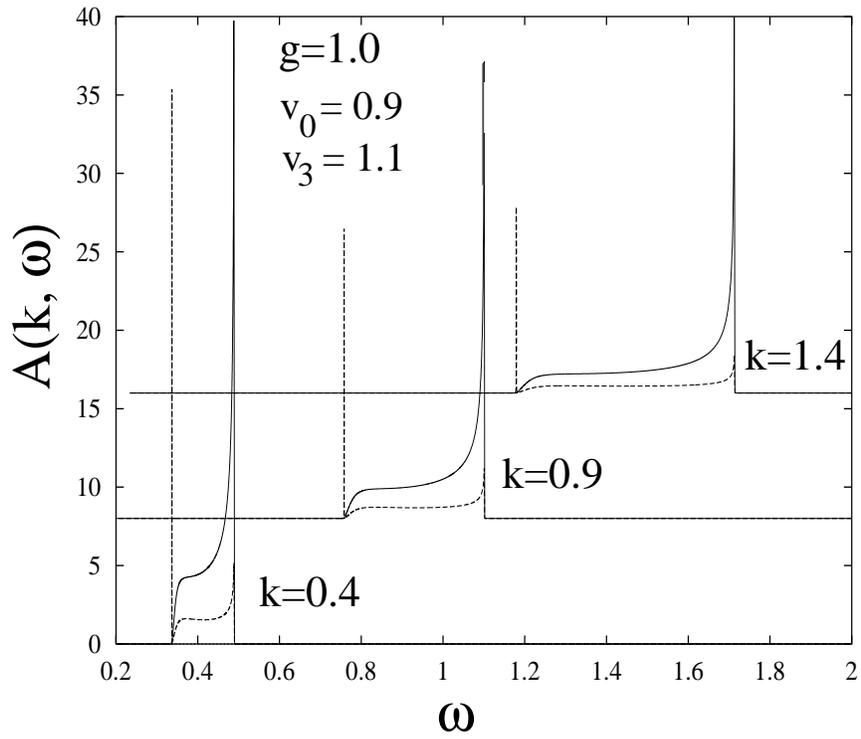}
\end{picture}
\vskip 0.4 truein
\caption{Spectral weight of the O(3) model. For clarity, we have shifted up
the curves for various momenta by 8 units. }
\end{center}
\end{figure}

\begin{figure}
\unitlength1.0cm
\begin{center}
\begin{picture}(7,6)
\epsfxsize=10.0cm
\epsfbox{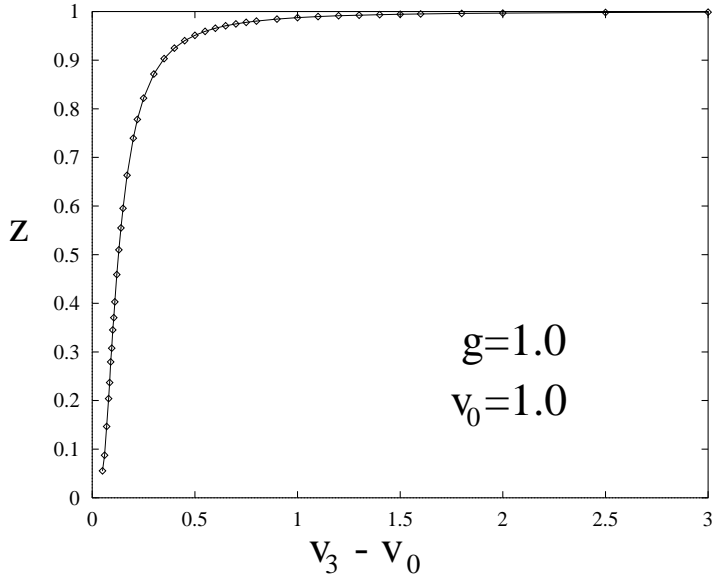}
\end{picture}
\vskip 0.4 truein
\caption{Quasi-particle weight $z$ of $\Psi^{(0)}$ in the $O(3)$ model. }
\end{center}
\end{figure}

\section{DISCUSSION and CONCLUSION}

In summary, we have demonstrated that in a system of interacting
chiral fermions with different velocities, due to restriction of the 
scattering phase space, the single Majorana fermion with
extremal velocity splits off from the Luttinger continuum to form
a sharp bound state or anti-bound state, leading to a system with two
qualitatively distinct spectral peaks and scattering rates. This is
a significant departure from the Luttinger liquid scenario and 
demonstrating a new class of 
of one-dimensional fixed point behavior.

In some sense, this new fixed point lies mid-way between the Luttinger
and the Fermi liquid, and is closest in character to the Marginal
Fermi liquid phenomenology originally introduced in the context of
cuprate metals.\cite{Varma} Unlike the Luttinger liquid, here we have a sharp
quasiparticle bound-state co-existing with an incoherent continuum.
Like the Luttinger liquid however, there are two velocities which
define the limits of the broad continuum  of excitations.
The qualitative picture of a sharp split-off peaks coexisting with broad
features in the spectral weights is a consequence of the
a phase space restriction relative to the Luttinger liquid.

Had we chosen to change two Majorana velocities at the same time,
so that $v_0=v_1$ and $v_2=v_3$, we would have reduced the symmetry still
further, to an $O(2) \times O(2)$ symmetry. In this case, it {\sl is}
possible to bosonize the Hamiltonian, obtaining the results
\bea
G_{3}(x,\tau)= \frac{1}{2\pi i x} \left[1- v_+ \tau/ix \right]^{-1/2+\gamma} 
\left[1-v_- \tau/ix \right]^{-1/2-\gamma}
\eea and $v_{\pm}= \left[v_0 + v_3 \pm \sqrt{(v_3 -v_0)^2 + (g/\pi)^2} \right]/2$, 
$\gamma = (v_3 -v_0)/(2 \sqrt{(v_3 -v_0)^2 + (g/\pi)^2})$. Curiously, this result may also be obtained by solving equations (9), even though, 
as far as we can see, the closed-loop cancellation 
is not sufficient in the case of the $O(2)\times O(2)$ model to cancel all
vertex corrections. This suggests
that a more general cancellation principle may be at
work, and that the range of validity of our solution may extend
beyond the $O(3)$ down to models with just $O(2)$ symmetry. At present
we have not been able to prove this result.  

Our work raises the question whether this kind of non-Fermi
Liquid behavior might survive in dimensions higher than one.  In
higher dimensions energy conservation and momentum conservation are
distinct constraints on scattering phase space, and for this reason
the Fermi surface reverts from a Luttinger to a Fermi liquid in higher
dimensions\cite{Metzner}.  By contrast, the O(3) model can not be
bosonized, and its unusual properties do not rely on the coincidence
between momentum and energy conservation. For this reason, there is at
least a grain of hope that this kind of behavior might be more robust
in higher dimensions.  We do in fact know that near infinite
dimensions\cite{HoColeman}, two lifetimes behavior persists in the
$O(3)$ model, but here, the thermodynamics near zero temperature is
that of a Fermi Liquid. The case of small, but finite dimensions is
however, still open.

{\em Acknowledgment}
We should like to thank Natan Andrei, Thierry Giamarchi, Alexei Tsvelik
and particularly Walter Metzner for discussions related to this work. 
This work was supported by NSF grant NSF DMR 96-14999.

\newpage

\end{document}